\documentclass[aps,prl,twocolumn]{revtex4-1}
\usepackage[latin1]{inputenc}
\usepackage{amsmath}
\usepackage{amsfonts}
\usepackage{amssymb}
\usepackage[dvipdfmx]{graphicx}
\usepackage{xcolor}
\usepackage{bm}
\usepackage[a4paper,top=2cm,bottom=2cm,left=2cm,right=2cm]{geometry}

\begin{document}
	
\author{Yuta Sakamoto}
\affiliation{Department of Physical Sciences, Aoyama Gakuin University, 5-10-1 Fuchinobe, Chuo-ku, Sagamihara, Japan}
\author{Takahiro Sakaue}
\thanks{corresponding author, sakaue@phys.aoyama.ac.jp}
\affiliation{Department of Physical Sciences, Aoyama Gakuin University, 5-10-1 Fuchinobe, Chuo-ku, Sagamihara, Japan}


\title{First passage time statistics of non-Markovian random walker: Onsager's regression hypothesis approach}
	
\begin{abstract}
First passage time plays a fundamental role in dynamical characterization of stochastic processes. Crucially, our current understanding on the problem is almost entirely relies on the theoretical formulations, which assume the processes under consideration are Markovian, despite abundant non-Markovian dynamics found in complex systems. Here we introduce a simple and physically appealing analytical framework to calculate the first passage time statistics of non-Markovian walkers grounded in a fundamental principle of nonequilibrium statistical physics that connects the fluctuations in stochastic system to the macroscopic law of relaxation. 
Pinpointing a crucial role of the memory in the first passage time statistics, our approach not only allows us to confirm the non-trivial scaling conjectures for fractional Brownian motion, but also provides a formula of the first passage time distribution in the entire time scale, and establish the quantitative description of the position probability distribution of non-Markovian walkers in the presence of absorbing boundary.
\end{abstract}

\maketitle

How long does it take for a random walker to reach a destination? Such a question on the first passage time (FPT) is relevant to a broad range of situations in science, technology and  every-day life applications as encountered, for instance, in diffusion-limited reactions~\cite{vanKampen_2007, Redner_2001,Szabo_1980}, barrier crossing~\cite{Kramers_1940,Hanggi_1990,Carlon_2018,Lavacchi_2022}, target search processes~\cite{Klafter_2007, Lomholt_2008}, cyclization of DNA molecule~\cite{Fixman_1974,Doi_1975,Sokolov_2003,Benichou_2015}, price fluctuation in market~\cite{Redner_2001} and spread of diseases~\cite{Lawley_2020}.
Today, the concept of the FPT and its importance in the study of stochastic processes are well recognized, and 
theoretical methods for its computation are standardized~\cite{vanKampen_2007, Redner_2001}. 
However, most of them are devised for Markovian random walkers, whose decision making does not depend on its past history, thus not applicable to non-Markovian walkers despite their ubiquitousness.

Indeed, a growing body of evidence suggests that the non-Markovian dynamics is found quite generally in rheologically complex matters typically, but not exclusively, with viscoelastic properties.
Classical examples are found in the diffusion of interacting particles in narrow channels~\cite{Wei_2000} and the motion of tagged monomers in long polymer chain~\cite{Panja_2010,Saito_2015}. Other notable representatives include colloidal particles in polymer solutions~\cite{Leibler_1996} or nematic solvents~\cite{Turiv_2013}, lipids molecules and cholesterols in cellular membrane~\cite{Jeon_2012}, proteins in crowded media~\cite{Bank_2005}, and chromosome loci~\cite{Kimura_2022} as well as membraneless organelles in living cells~\cite{Benelli_2021}.
Such systems commonly exhibit a slow dynamics in the form of sub-diffusion ${\rm MSD} (t) \sim t^{\alpha}$ characterized by the anomalous exponent  $\alpha < 1$, where ${\rm MSD}(t)$ stands for the mean-square displacement of the observed particle during the time scale $t$ . Here the physical mechanism at work is the interaction of observed degree of freedom with the collective modes with broad range of time scales underlying complex environment. Because of its importance in e.g. intracellular transport, the theoretical tools to describe/diagnose such anomalous diffusion phenomenology have been well developed in the last few decades~\cite{Metzler_2014}. However, most of them rely on MSD and related quantities, while much less attention has been paid to the FPT, despite its fundamental and practical importance to characterize the underlying stochastic process. This is particularly true for systems possessing {\it memory}, as nontrivial information on the history dependence of the system is encoded in the FPT statistics~\cite{Bray_2013}. It has long been known that the anomalous transport properties affect the rates of chemical and biochemical reactions~\cite{Minton_2001}, and such reactions are initiated by the encounter of reactant molecules, so precisely quantified by means of the FTP statistics. 

\begin{figure}[b]
	\centering
	\includegraphics[width=0.45\textwidth]{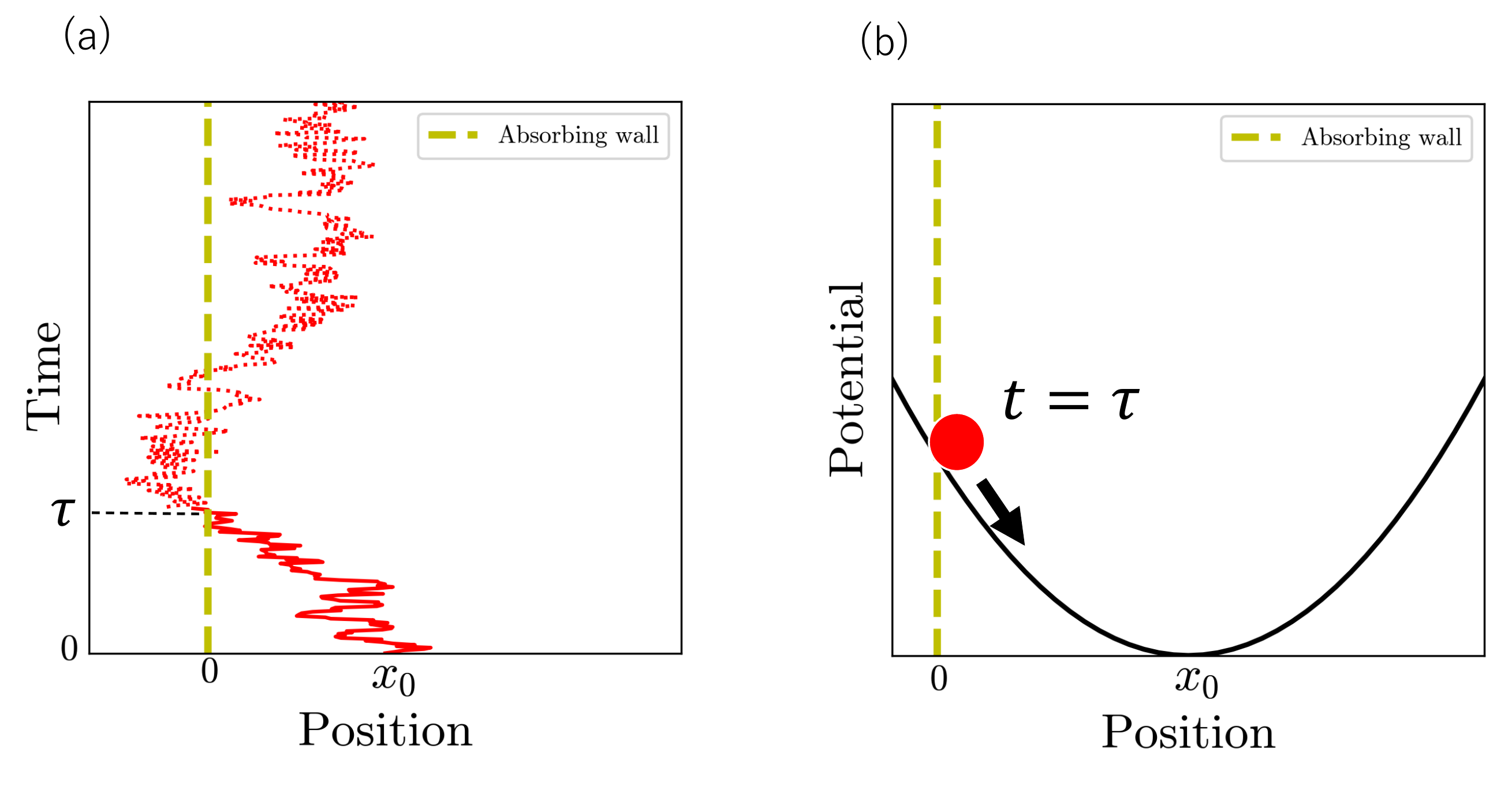}
	\caption{Regression hypothesis applied to non-Markovian walkers.  (a) Example trajectory of fBM with $\alpha=0.5$ starting from the initial position $x=x_0$. Before (after) the first hitting on absorbing boundary at $x=0$, the trajectory is drawn by solid (dotted) curve. First passage event can be viewed as a large fluctuation to create a non-equilibrium state at $t=\tau$.  (b) After the first passage ($t > \tau$), the process follows, on average, the macroscopic relaxation law, for sub-diffusive fBM, represented by the harmonic restoring force, whose spring constant gets smaller algebraically in longer time scales.}
	\label{fig1}
			\vspace{0.2 cm}
\end{figure}

Unfortunately, our current understanding on the FPT of non-Markovian walker lags far behind that of Markovian counterpart, where the difficulty is largely associated to the lack of appropriate theoretical foothold~\cite{Amitai_2010,Bray_2013, Guerin_2016}. While the Fokker-Planck equation and its related methods play a key role to analyze the time evolution of the probability distribution of the Markovian walkers, their careless application is problematic for walkers with memory, a defining property of the non-Markovian process. 
At present, available results are quite limited with notable examples being the perturbative and scaling arguments to estimate the asymptotic exponents characterizing the distribution of FPT and related quantities in unbounded domain~\cite{Krug_1997,Bray_2013,Majumdar_2009,Majumdar_2011}, some approximation schemes to calculate the mean FPT of polymer looping process~\cite{Szabo_1980,Fixman_1974,Doi_1975,Sokolov_2003,Benichou_2015}, and more recent analytical treatment to compute the mean FPT in confined domains~\cite{Guerin_2016}. However, neither of the full distribution of FPT or position distribution of non-Markovian walkers in the presence of boundary are available, making the computation of these quantities in non-Markovian processes fundamental challenge. 


In this Letter, we provide a simple and physically appealing method to calculate the FPT statistics of non-Markovian walkers by identifying the moment of first passage ($t = \tau$) as an initial condition for the relaxation process afterwards ($ t > \tau$), see Fig. 1. Our argument is thus rooted in
 a non-Markovian extension of the {\it regression hypothesis} of Onsager, a corner stone for the development in the nonequilibrium statistical physics~\cite{Onsager_1931}. We obtain an exact integral equation for the FPT distribution, the analysis of which yields, in addition to its asymptotic decay exponent, full functional form in leading order over entire time scales, and also the walker's probability distribution function. Importantly, our formalism allows one to unveil how and why the textbook standard ``method of image''~\cite{Chandrasekhar_1943,Redner_2001} breaks down by pinpointing the role of memory built up during the first passage process. Here we focus on the sub-diffusive fractional Brownian motion~\cite{Mandelbrot_1968} (fBM with $\alpha < 1$), an important class of non-Markovian walkers found in widespread complex systems including living cells and nuclei~\cite{Kimura_2022,Jeon_2012,Bank_2005,Benelli_2021}. 


\begin{figure}[ht]
	\centering
	\includegraphics[width=0.38\textwidth]{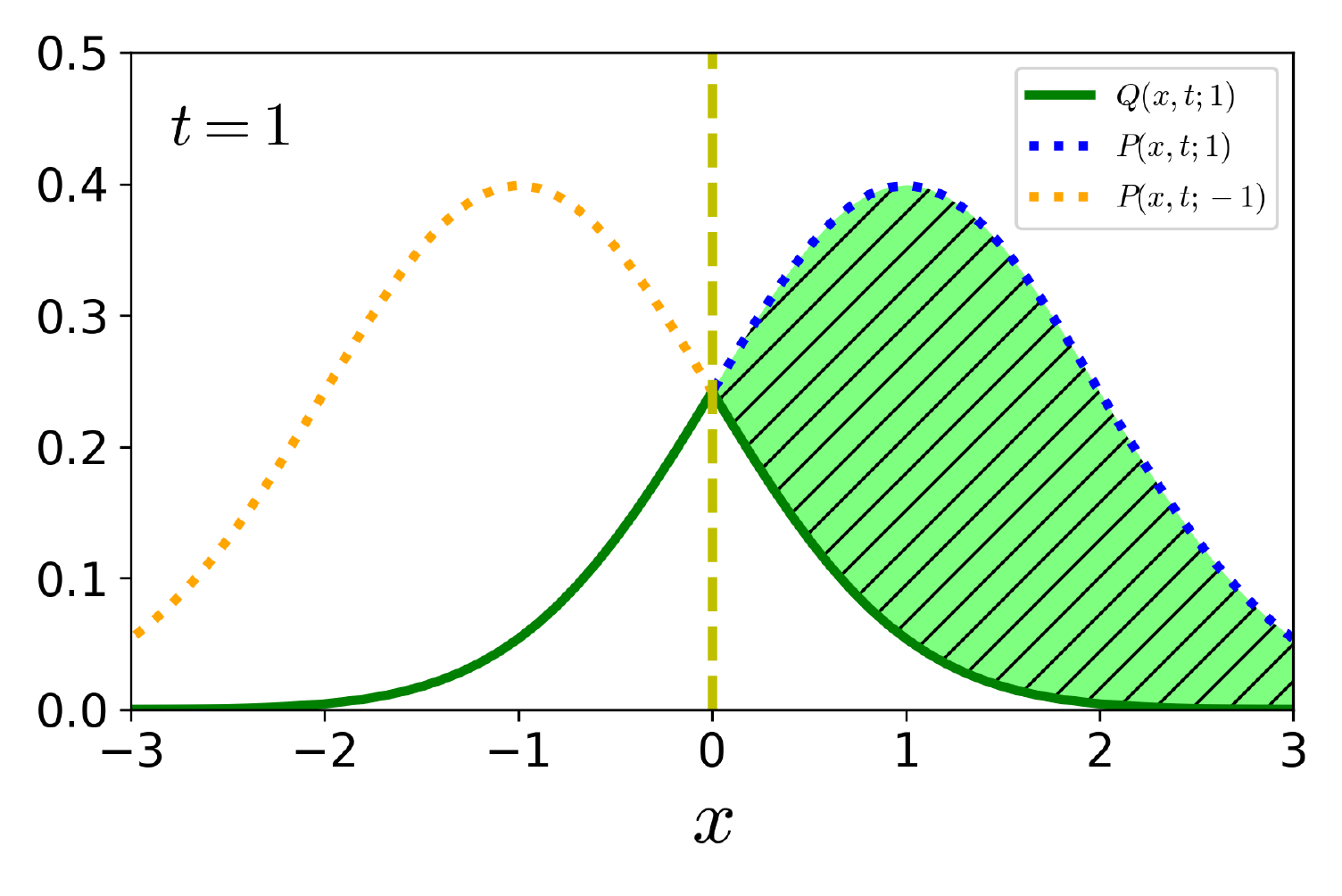}
	\caption{Illustration of the method of image. For Markovian walkers ($\alpha=1$), $Q(x,t;1)$ can be constructed by the method of image. Integrating  Eqs.~(\ref{P_P0_Q}) over the entire space (including negative domain), one finds $S(t;1)= 1-\int_{-\infty}^{\infty}Q(x,t;1) dx$, where the surviving probability $S(t;1)= \int_0^{\infty} P_{+}(x,t;1) dx$ is denoted by the hatched area. Equivalent to the above relation is $\int_{0}^{\infty}Q(x,t;1) dx= (1-S(t;1))/2$ thanks to the reversal symmetry of $Q(x,t;1)$ with respect to $x=0$, producing a factor $1/2$. The same relation is obtained by integrating Eq.~(\ref{Q}) over the positive $x$ domain with $\langle x(t) \rangle_{{\rm FPT}=\tau}=0$. }
	\label{fig2}
			\vspace{0.0 cm}
\end{figure}

{\it Generalized Langevin equation and power-law memory kernel} \ --
As a paradigm, consider a random walker in one dimensional half space with an absorbing boundary at origin. 
A walker is initially positioned at $x=x_0 (> 0)$ at $t=0$, and evolves according to the following generalized Langevin equation:
\begin{eqnarray}
\frac{dx(t)}{dt} = \int_0^t \mu(t-t') f(t') dt'  + \eta(t) 
\label{GLE}
\end{eqnarray}
where $f(t)$ and $\eta(t)$ are, respectively, a time-dependent external force and the noise acting on the walker~\cite{Saito_2015}. The latter is assumed to be Gaussian with zero mean and its auto-correlation is related to the mobility kernel via the fluctuation-dissipation relation $\langle \eta(t) \eta(t') \rangle = T \mu(|t-t'|)$ with $T$ being the noise strength. The memory effect is encoded in $\mu(t)$, for which we assume for  large $t$ the power-law decay $\mu(t) \simeq - T^{-1} {\mathcal D}_{\alpha} t^{\alpha-2}$ ($ 0 < \alpha <1$) in addition to instantaneous response $\mu(t) = 2 \gamma^{-1} \delta(t)$ at short time, where $\gamma$ is a bare friction coefficient. Finally, we require on physical ground $\int_0^{\infty}\mu(t) dt =0$ such that Eq.~(\ref{GLE}) describes the sub-diffusive fBM with the MSD exponent $\alpha$. This {\it sum rule} is a consequence of the relaxation nature of the sub-diffusive fBM, which is caused by the visco-elastic effect~\cite{Saito_2015}.
 For a free walker ($f=0$) in free space (no boundrary), its position probability distribution $P(x,t;x_0) $ is simply given by  ${\mathcal N}(x,x_0,2 {\mathcal D}_{\alpha}t^{\alpha})$, where ${\mathcal N}(x,A,B) = (2 \pi B)^{-1/2}e^{(x-A)^2/2B}$ denotes Gaussian distribution of $x$ with the average $A$ and the variance $B$.
 
{\it Process after first-passage} \ --
We now set a stage by introducing an absorbing boundary at the origin $x=0$ such that the walker performs fBM in half space $x>0$ with the same initial condition as before. Using the free space propagator $P(x,t;x_0)$, the walker's position probability $P_{+}(x,t; x_0)$ is now represented as
\begin{eqnarray}
P_{+}(x,t; x_0) = P(x,t;x_0) - Q(x,t;x_0)  
\label{P_P0_Q}
\end{eqnarray}
where $Q(x,t;x_0)$ is the position distribution of {\it dead walker}, who touched the absorbing boundary by this moment. 
Note that while one usually looks at the walker's behavior in physical domain ($x \ge 0$) up to the absorption ($t \le \tau$) in the context of FPT,  Eq.~(\ref{P_P0_Q}) holds in entire space and time domains in a sprit similar to ~\cite{Guerin_2016}; the absorbing boundary at $x=0$ necessitates $P(x,t;x_0) = Q(x,t;x_0)$ for $x \le 0$.
Using the FPT distribution $F(\tau; x_0)$,  $Q(x,t;x_0)$ is represented  as
\begin{eqnarray}
Q(x,t;x_0) = \int_0^t F(\tau; x_0) \ P(x,t ;x_0| {\rm {\scriptstyle FPT}} = \tau) d \tau
\label{Q}
\end{eqnarray}
where $P(x,t;x_0 | {\rm {\scriptstyle FPT}} = \tau)$ is the conditional probability of the walker's position at time $t$ after its first passage at time $\tau$. Being the Gaussian process, one expects the form 
\begin{eqnarray}
P(x,t;x_0 | {\rm {\scriptstyle FPT}} = \tau) = {\mathcal N}(x, \langle x(t) \rangle_{\rm { FPT}=\tau}, 2 {\mathcal D}_{\alpha}(t - \tau)^{\alpha}) \nonumber \\
 .
\label{P_x0_FPT}
\end{eqnarray}
In the absence of memory effect, $\langle x(t) \rangle_{{\rm FPT}=\tau}=0$ irrespective of the starting position $x_0$. Then, by noting $\int_0^t  F(t'; x_0) dt' = 1 - S(t;x_0)$,  integrating Eq.~(\ref{P_P0_Q}) over half space leads to a classical result of the survival probability $S(t;x_0) \equiv \int_0^{\infty}  P_{+}(x,t;x_0) dx= {\rm erf}(x_0/\sqrt{4 {\mathcal D}_1 t})$ for Markovian case, see Fig.~2. Although not applicable to non-Markovian walker, the above calculation highlights $\langle x(t) \rangle_{{\rm FPT}=\tau}$, which generally depends on $x_0$, as a central quantity to account for the memory effect in the first passage statistics.

{\it History-dependent relaxation:  regression hypothesis view} \ --
A key idea to quantify  $\langle x(t) \rangle_{{\rm FPT}=\tau}$ comes from the fundamental connection between fluctuation and response in nonequilibrium statistical physics.
In his seminal paper, Onsager pointed out that the decay of mesoscopic fluctuations follow, on average, the macroscopic law of relaxation~\cite{Onsager_1931}.
Applying this so-called {\it regression hypothesis} to our problem,  we view the process after the first passage $t > \tau$ as a relaxation process, whose ``initial'' condition $x(\tau)=0$ can be prepared either naturally (by fluctuation) or artificially (by external force), see Fig.~1. In the latter scenario, we take the sub-ensemble of walkers whose FPT is $\tau$, and describe their average time evolution using Eq.~(\ref{GLE}) with the constant force $f(t) = f_0$ for $t<\tau$.
This leads to
\begin{eqnarray}
\langle {\dot x}(t) \rangle_{{\rm FPT}=\tau} = f_0 \int_0^t  \mu(t')  dt' \qquad (t < \tau)
\end{eqnarray}
then, identifying $\langle {\dot x}(\tau) \rangle_{{\rm FPT}=\tau}  \simeq - x_0/\tau$, we find
\begin{eqnarray}
f_0 \simeq - \frac{T x_0}{{\mathcal D}_{\alpha}}  \tau^{-\alpha} .
\label{f0}
\end{eqnarray}
Now the desired non-equilibrium state is prepared at $t=\tau$, at which we switch off the force. The subsequent relaxation is described, again using Eq.~(\ref{GLE}), by
\begin{eqnarray}
\langle {\dot x}(t) \rangle_{{\rm FPT}=\tau} = f_0  \int_{t-\tau}^{t}  \mu (t') dt' ,    \qquad (t > \tau)
\end{eqnarray}
whose integral with respect to time leads to $\langle x(t) \rangle_{{\rm FPT}=\tau}$, where a numerical coefficient implicit in Eq.~(\ref{f0}) is fixed by requiring $\langle x(t) \rangle_{{\rm FPT}=\tau} \rightarrow x_0$ for  $t /\tau \gg 1$ as a consequence of the sum rule.
Collecting all together, our analytical formulation is summarized as the following integral equation~\cite{SI}:
\begin{eqnarray}
1- \mathrm{erf} \left( \frac{1}{\sqrt{2 t^{\alpha}}}\right) = \int_0^t  F(\tau;1)\left[ 1- \mathrm{erf}(h(t,\tau) ) \right]  d\tau \nonumber \\
 \label{Integ_eq_F}
\end{eqnarray}
with the memory function
\begin{eqnarray}
h(t,\tau) = \frac{1}{\sqrt{2 (t - \tau)^{\alpha} }}\left[ 1 + \left( \frac{t}{\tau} - 1 \right)^{\alpha} - \left( \frac{t}{\tau}\right)^{\alpha} \right] . 
\end{eqnarray}
From here onwards, we measure the length and the time in unit of $x_0$ and $\tau_{x_0}=(x_0^2/2 {\mathcal D}_{\alpha})^{1/\alpha}$, respectively, which are the sole characteristic length and time scales in the problem, making the initial position $x_0=1$ upon rescaling.

{\it First passage time distribution} \ --
We now determine the leading order solution of Eq.~(\ref{Integ_eq_F}) in the form
\begin{eqnarray}
F(\tau;1) = {\mathcal C}_{\alpha} \exp{\left[-\left( \frac{1}{2 \tau}\right)^{\omega} \right]} \tau^{-(1+p)}
\label{F_ansatz}
\end{eqnarray}
where ${\mathcal C}_{\alpha}$ is a normalization constant. 
This function, a generalization of the Markovian result~\cite{Redner_2001} $\omega=1$, $p=1/2$, exhibits a peak at $\tau = \tau^*= (1/2)(\omega / (1+p))^{1/\omega}$ and develops a power-law tail $F(\tau;1) \sim \tau^{-(1+p)}$ at $ \tau \gg \tau^*$. 
With this in mind, we plug the ansatz ~(\ref{F_ansatz}) into Eq.~(\ref{Integ_eq_F}) and perform the asymptotic analysis, which yields $p=1- \alpha/2$ in agreement with previous scaling argument~\cite{Krug_1997,Bray_2013}. In addition, our formulation allows us to obtain the exponent $\omega$, which satisfies the relation
\begin{eqnarray}
\frac{(2-\alpha)^{2 \omega} (2 + \alpha)^{\alpha}}{(2 \omega)^{\alpha}} = \left( \frac{3}{2}\right)^{\omega} c_1^{\omega(\alpha-1)}
\label{omega}
\end{eqnarray}
with a numerical constant $c_1$ of order unity~\cite{SI}.

\begin{figure}
	\centering
	\includegraphics[width=0.5\textwidth]{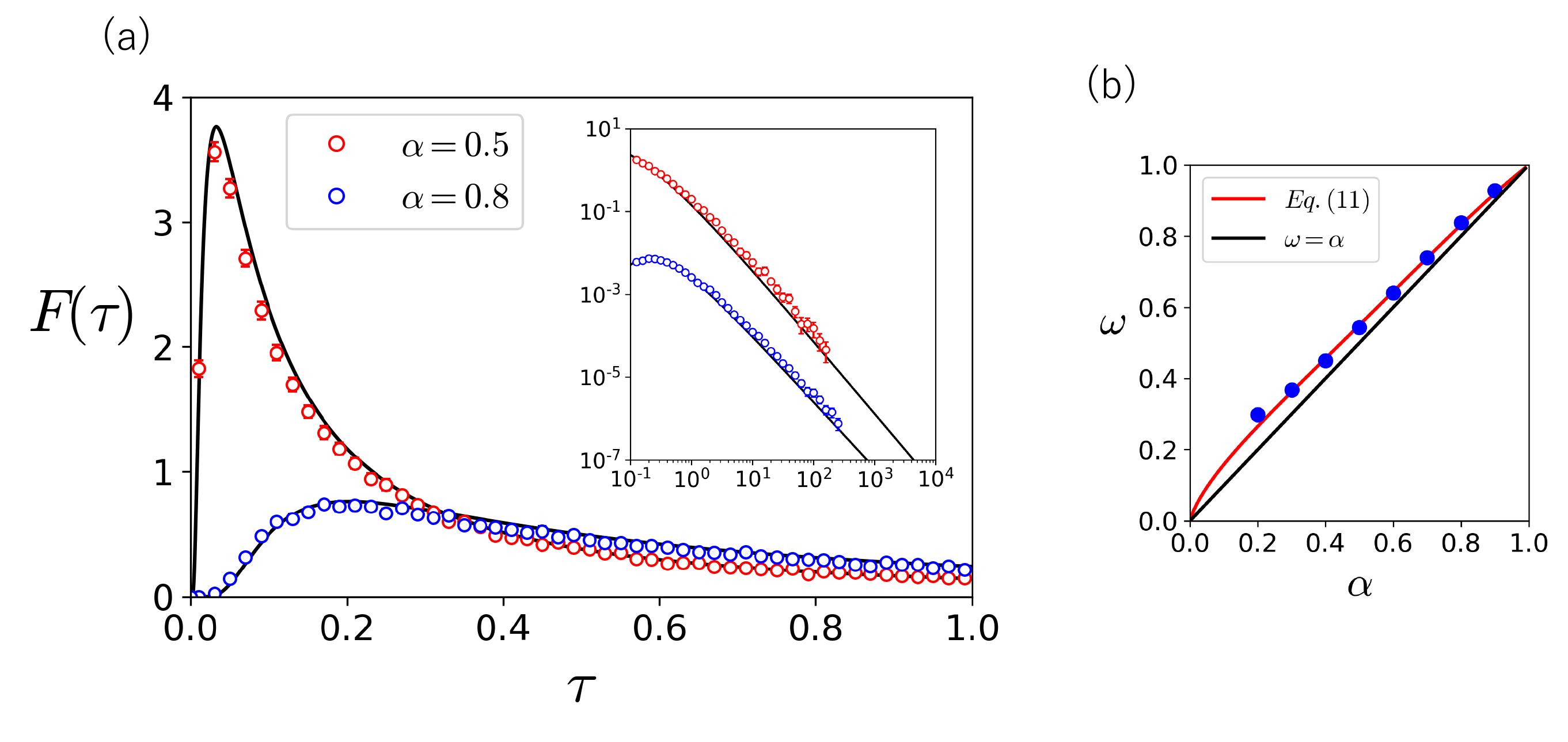}
	\caption{ FPT distribution of non-Markovian walkers. (a) FPT distribution $F(\tau; 1)$ for sub-diffusive fBM ($\alpha=0.8, 0.5$).  Inset shows the double logarithmic plot of large $\tau$ regime, where the asymptotic slope $p+1 = 2-\alpha/2$ is clearly visible. The data for $\alpha=0.8$ is shifted downward ($\times 10^{-2}$) for visual clarity. Both in main panel and inset, symbols represent simulation results and the curves correspond to the analytical formula~(\ref{F_ansatz}) with $p=1-\alpha/2$ and $\omega$ given by Eq.~(\ref{omega}). The error bars represent 95 \% CI. 
	(b) Exponent $\omega$ as a function of $\alpha$, which characterizes the early time regime in FPT distribution. Blue solid circles are obtained by fitting the numerical simulation data for several $\alpha$ values (two of them shown in Fig. 2(a)) with the formula ~(\ref{F_ansatz}). Fitting these data with Eq.~(\ref{omega}) fixes the parameter $c_1 = 0.1$.}
	\label{fig3}
			\vspace{0.2 cm}
\end{figure}

\begin{figure*}[]
	\centering
	\includegraphics[width=0.8\textwidth]{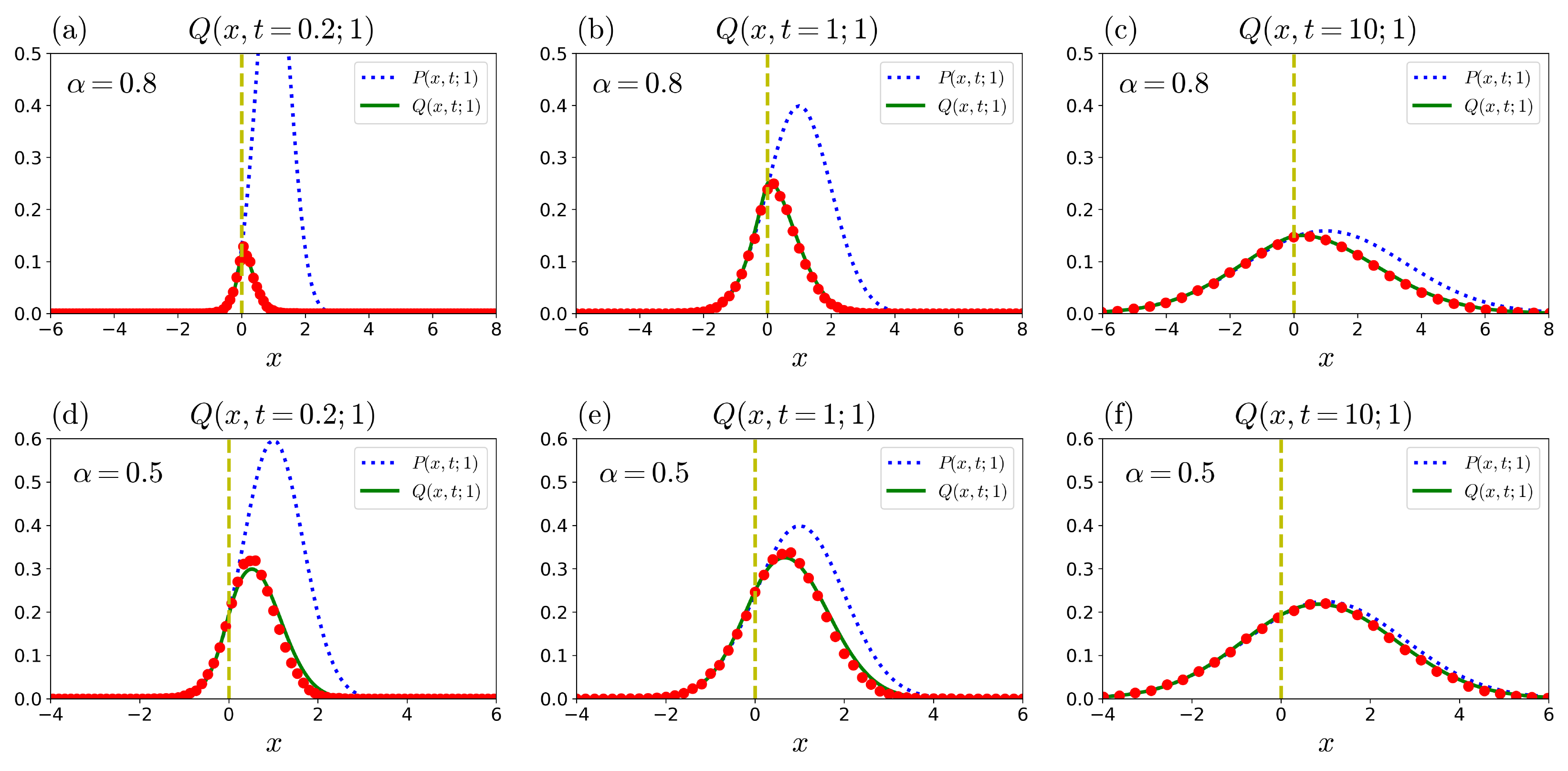}
	\caption{ Probability distribution $Q(x,t;1)$ of the position of absorbed sub-diffusive walkers. Plots of $Q(x,t;1)$ for sub-diffusive fBM (a)-(c) with $\alpha=0.8$  and (d)-(f) with $\alpha=0.5$  at early, middle and late times ($t =0.2,\ 1,\  10$, respectively). Analytical prediction (green solid curve) is obtained using Eqs.~(\ref{Q}),~(\ref{P_x0_FPT}) and~(\ref{F_ansatz}), which quantitatively reproduces the numerical simulation results (red circles). The error bar evaluated as 95 \% CI is smaller than the size of symbol.  Blue dashed curve represent the free space propagator $P(x,t;1)$. The asymmetry in $Q(x,t;1)$ grows with the memory effect, which is stronger for smaller $\alpha$. }
	\label{fig4}
			\vspace{0. cm}
\end{figure*}

In Fig.~3 , we compare our analytical formula for $F(\tau;1)$ with the results obtained from numerical simulation~\cite{SI}.
As shown, the agreement is excellent, encompassing the short time singularity to the peak, and the eventual long time power-law tail, which are characterized by the exponents $\omega$ and $p$, respectively. The peak position $\tau^*$ is rather sensitive to the value of $\omega$. This is particularly true for small $\omega$, which is the case for the small $\alpha$, shifting the peak position $\tau^*$ vanishingly small in the limit $\alpha \rightarrow 0$.

\begin{figure}[h]
	\centering
	\includegraphics[width=0.35\textwidth]{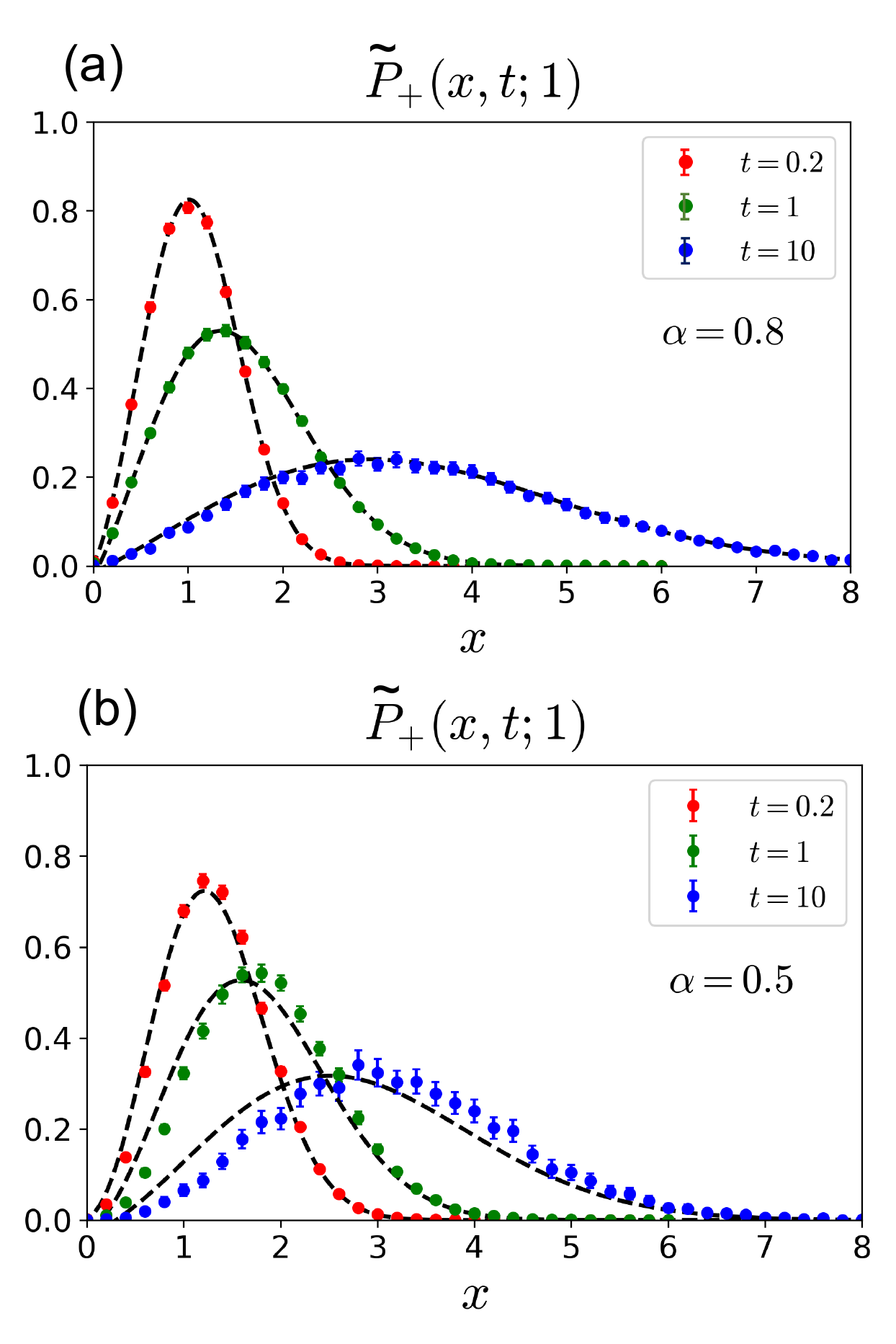}
	\caption{Probability distribution $P_{+}(x,t;1)$ of the position of survived sub-diffusive walkers. Plots of the normalized position probability ${\tilde P}_{+}(x,t;1) \equiv P_{+}(x,t;1)/S(x;1)$ for sub-diffusive fBM with (a) $\alpha=0.8$ and (b) $\alpha=0.5$ at early, middle and late times ($t =0.2,\ 1,\  10$, respectively). Analytical prediction (dashed curve) is obtained using Eq.~(\ref{P_P0_Q}), which reproduces the numerical simulation results (symbols). Error bars represent 95 \% CI.}
	\label{fig5}
			\vspace{0.1 cm}
\end{figure}

{\it Probability distributions of dead and survived walkers} \ --
We are now in a position to take a close look at $Q(x,t;1)$ that is the distribution of walkers after their first passage. From Eqs.~(\ref{Q}) and~(\ref{P_x0_FPT}), we immediately find the memory effect in the form of restoring force represented by nonzero $\langle x(t) \rangle_{{\rm FPT} = \tau}$ breaks the reversal symmetry with respect to $x=0$, i.e., $Q(x,t;1) \neq Q(-x,t;1)$ that clearly manifests the breakdown of the image method (Figs.~2, 4)~\cite{SI}. The  value of $\langle x(t) \rangle_{{\rm FPT} = \tau}$ corresponds to the peak position of $P(x,t ;1| {\rm {\scriptstyle FPT}} = \tau)$, which is zero initially ($t=\tau$), and slowly evolves with time towards $x=x_0$. Such a distribution $P(x,t ;1| {\rm {\scriptstyle FPT}} = \tau)$ characterizes the subensemble of walkers with fixed FPT, whose superimposition with the weight $F(\tau;1)$ results in $Q(x,t;1)$, see Eq.~(\ref{Q}). As Fig.~4 shows, our analytical prediction of $Q(x,t;1)$ quantitatively captures the results obtained by numerical simulations.

In Fig.~5, we plot the normalized position probability ${\tilde P}_{+}(x,t;1) \equiv P_{+}(x,t;1)/S(x;1)$ of the survival walker from Eq.~(\ref{P_P0_Q}). Again, our prediction captures all the salient features seen in numerical simulations. One notable feature here is that the slope $(\partial {\tilde P}_{+}(x,t;1)/\partial x)_{x \rightarrow 0}$ at the boundary is vanishingly small~\cite{Kantor_2007}. 
Such an anomalous behavior of ${\tilde P}_{+}(x,t;1) \sim x^{\delta}$ close to the boundary with non-trivial exponent $\delta$ can be quantified from our expression for $Q(x,t;1)$ as follows. Note first that in long time limit $t \gg 1$ ( $\Leftrightarrow  \ x_0^2/ {\mathcal D}_{\alpha}t^{\alpha} \ll 1$ in original unit), the asymptotic behavior of ${\tilde P}_{+}(x,t;1)$ is obtained by taking $x_0 \rightarrow 0$ limit~\cite{Majumdar_2009}. For the walker absorbed at time $\tau$, its characteristic travel distance during the subsequent time interval $s = t-\tau$ is evaluated as $\Delta x(s) \sim s^{\alpha/2}$.
This indicates that, for a given location $x$, the walker only starts substantially contributing to $Q(x,t;1)$ after the time $t^{(x)} = x^{2/\alpha}$. From Eq. ~(\ref{Q}), we thus find
\begin{eqnarray}
Q(x,t;1) &\sim& \int_{t^{(x)}}^{t-\tau^*} (t-s)^{-(2-\alpha/2)} \ s^{-\alpha/2} \ ds   \nonumber \\ 
&\sim& t^{-\alpha/2} \left( 1 - t^{-(2-\alpha)} x^{(2-\alpha)/\alpha}\right)
\end{eqnarray}
The first term cancels the free space propagator $P(x,t;1) \sim t^{-\alpha/2}$, leaving $P_{+}(x,t;1) \sim t^{-(2-\alpha/2)} x^{(2-\alpha)/\alpha}$, or equivalently, ${\tilde P}_{+}(x,t;1) \sim t^{-1} x^{(2-\alpha)/\alpha}$. The predicted exponent $\delta =  (2-\alpha)/\alpha$ agrees with that obtained from heuristic scaling argument~\cite{Majumdar_2009}.

For the Markovian case $\alpha =1$, the slope at the boundary is finite ($\delta =1$), which multiplied by diffusion coefficient is the outgoing flux. The peculiar nature of the flux for $\alpha \neq 1$ case implies the breakdown of the Fick's law, and makes the implementation of a reflective boundary non-trivial. This rephrases a fact that there is no diffusion (more generally Fokker-Planck) equation for non-Markovian walkers in the ordinary sense. 

In conclusion, we have provided a natural framework with which the first passage process of non-Markovian walkers can be analyzed. It is very simple, yet has a quantitative predictability as we have demonstrated here for the system with persistent memory, i.e., sub-diffusive fBM.  We expect that the proposed method with suitable extension and generalization
will find versatile applicability to explore rich  FPT problems in
non-Markovian processes.

\subsection*{Acknowledgements}
We thank E. Carlon for fruitful discussion. This work is supported by JSPS KAKANHI (Grants No. JP18H05529 and JP21H05759).

\bibliography{stochastic}

\end{document}


\author{Yuta Sakamoto}
\affiliation{Department of Physical Sciences, Aoyama Gakuin University, 5-10-1 Fuchinobe, Chuo-ku, Sagamihara, Japan}
\author{Takahiro Sakaue}
\thanks{corresponding author, sakaue@phys.aoyama.ac.jp}
\affiliation{Department of Physical Sciences, Aoyama Gakuin University, 5-10-1 Fuchinobe, Chuo-ku, Sagamihara, Japan}


\title{Supplementary Material}

\maketitle

\section{ Derivation of integral equation} 
We start with Eq.~(2) in the main text;
\begin{eqnarray}
P_{+}(x,t; 1) = P(x,t;1) - Q(x,t;1)  
\label{P+_P_Q}
\end{eqnarray}
Here the walker's initial position $x_0 >0$ is a sole length scale in the problem, and we measure the length in unit of $x_0$. Similarly, we introduce the unit of time $\tau_{x_0}=(x_0^2/2 {\mathcal D}_{\alpha})^{1/\alpha}$, which corresponds to the time scale for a walker to diffuse over the length scale $x_0$. Note the rescaled initial position $x_0=1$, and
\begin{eqnarray}
P(x,t;1)& =& \frac{1}{\sqrt{2 \pi t^{\alpha}}} e^{-\frac{(x-1)^2}{2 t^{\alpha}}} \\
Q(x,t;1) &=& \int_0^t F(\tau; 1) \ P(x,t ;1| {\rm {\scriptstyle FPT}} = \tau) d \tau \nonumber \\
&=&  \int_0^t F(\tau; 1)   \frac{1}{\sqrt{2 \pi (t-\tau)^{\alpha}}} e^{-\frac{\{x- \langle x(t) \rangle_{\rm { FPT}=\tau}\}^2}{2 (t-\tau)^{\alpha}}}   d \tau 
\label{P_Q}
\end{eqnarray}
where
\begin{eqnarray}
 \langle x(t) \rangle_{\rm { FPT}=\tau} = 1 + \left(\frac{t}{\tau} -1\right)^{\alpha} - \left( \frac{t}{\tau}\right)^{\alpha}  \qquad (t \ge \tau)
 \label{u}
 \end{eqnarray}
 is the average trajectory of the walkers after the first-passage at $t=\tau$, which is calculated by applying the regression hypothesis idea of Onsager as explained in the main text.

The integral of Eq.~(\ref{P+_P_Q}) over the half space $(x \ge 0)$ leads to
\begin{eqnarray}
S(t;1) = \frac{1}{2}\left\{ \mathrm{erf}\left( \frac{1}{\sqrt{2 t^{\alpha}}} \right)  + 1\right\} - \frac{1}{2}\int_0^t F(\tau;1) \ \mathrm{erf}\left( \frac{\langle x(t) \rangle_{\rm { FPT}=\tau}}{\sqrt{2 (t-\tau)^{\alpha}}} \right)  d\tau
\end{eqnarray}
where $S(t;1)$ is the survival probability. Noting the relation $S(t;1) = 1 - \int_0^t F(\tau;1) d\tau$, the above equation is transformed to
\begin{eqnarray}
1- \mathrm{erf} \left( \frac{1}{\sqrt{2 t^{\alpha}}}\right) = \int_0^t  F(\tau;1)\left[ 1- \mathrm{erf}(h(t,\tau) ) \right]  d\tau \label{Integ_eq_F}
\end{eqnarray}
with the memory function $h(t,\tau) =  \frac{\langle x(t) \rangle_{\rm { FPT}=\tau}}{\sqrt{2 (t-\tau)^{\alpha}}}$, which is an exact integral equation to determine $F(\tau, 1)$  (Eq.~(8) in the main text).

\section{ Analysis of integral equation}
To analyze the integral equation~(\ref{Integ_eq_F}), we first rewrite the memory function as
\begin{eqnarray}
h(t,\tau) =  \frac{t^{-\alpha/2}}{\sqrt{2}}  g(u)
\end{eqnarray}
with
\begin{eqnarray}
g(u) &=& (1-u)^{-\alpha/2}(1-u^{-\alpha}) + (1-u)^{\alpha/2}u^{-\alpha}
\end{eqnarray}
where $u \equiv \tau/t$.
The error function in the integrand is expanded as
\begin{eqnarray}
\mathrm{erf} (h(t,\tau)) =  \mathrm{erf} \left(  \frac{t^{-\alpha/2}}{\sqrt{2}} \right) + \sqrt{ \frac{2}{\pi}} t^{-\alpha/2}(g(u)-1) + \mathcal{O}(t^{-3 \alpha/2})
\end{eqnarray}
Neglecting higher order terms $\mathcal{O}(t^{-3 \alpha/2})$, Eq.~(\ref{Integ_eq_F}) becomes
\begin{eqnarray}
S(t;1) \left[ 1 -  \mathrm{erf}  \left(  \frac{t^{-\alpha/2}}{\sqrt{2}} \right) \right] \simeq \sqrt{\frac{2}{\pi}} \ t^{1- \alpha/2}\int_0^1 F(\tau(u);1)\left\{ 1 - g(u) \right\} du
\label{Integ_eq_F2}
\end{eqnarray}

Motivated by the known analytical solution 
\begin{eqnarray}
F(\tau;1) = {\mathcal C}_{1} \exp{\left[-\left( \frac{1}{2 \tau}\right) \right]} \tau^{-3/2}
\end{eqnarray}
for the Markovian case ($\alpha=1$), where ${\mathcal C}_{1}$ is a normalization constant, we seek for the solution in the form
\begin{eqnarray}
F(\tau;1) &=& {\mathcal C}_{\alpha} \exp{\left[-\left( \frac{1}{2 \tau}\right)^{\omega} \right]} \tau^{-(1+p)} \nonumber \\
&=&{\mathcal C}_{\alpha} t^{-(1+p)}  \exp{\left[-\left( \frac{1}{2 tu}\right)^{\omega} \right]}  u^{-(1+p)} 
\label{F_ansatz}
\end{eqnarray}
Substituting the above ansatz in Eq.~(\ref{Integ_eq_F2}), we obtain
\begin{eqnarray}
&&S(t;1) \left[ 1 -  \mathrm{erf}  \left(  \frac{t^{-\alpha/2}}{\sqrt{2}} \right) \right]  \nonumber \\
&&\simeq \sqrt{\frac{2}{\pi}}  {\mathcal C}_{\alpha} \ t^{-(p+\alpha/2)}   \int_0^1 e^{-\left( \frac{1}{2 tu}\right)^{\omega} } \left[  \alpha  u^{-(\alpha + p)}(1 + {\mathcal O}(u)) - \frac{\alpha}{2}u^{-p} (1 + {\mathcal O}(u) )\right] du \nonumber \\
\label{Integ_eq_F3}
\end{eqnarray}
To evaluate the above integral, we note the following:
\begin{eqnarray}
\int_0^1  e^{-\left( \frac{1}{2 tu}\right)^{\omega} } u^{-\kappa}  du  &\simeq&  \int_{u^*}^1  u^{-\kappa} du 
\end{eqnarray}
where $u^* = c_1 t^{-1} (\omega/\kappa)^{1/\omega}$ with $c_1$ being a numerical constant of order unity.

Then, at leading order in $1/t$, Eq.~(\ref{Integ_eq_F3}) becomes
\begin{eqnarray}
S(t;1) \simeq \sqrt{\frac{2}{\pi}}  \  {\mathcal C}_{\alpha}  t^{-(1-\alpha/2 )} \frac{\alpha}{\alpha+p-1} \left( c_1 \left( \frac{\omega}{\alpha +p}\right)^{1/\omega}\right)^{1-\alpha -p}
\end{eqnarray}
which is asymptotically correct at large $t$.
Calculating $ - dS(t;1)/dt$ and comparing it with the assumed form of $F(t;1)$, we find the persistence exponent
\begin{eqnarray}
p=1 - \frac{\alpha}{2}
\end{eqnarray}
in agreement with earlier scaling argument~\cite{Krug_1997}.
In addition, by comparing two expressions of {\it prefactor}, we find a relation between $\omega$ and $\alpha$;
\begin{eqnarray}
(2-\alpha) \left( \frac{2+\alpha}{2 \omega}\right)^{\alpha/(2\omega)} c_1^{-\alpha/2} = c_2
\end{eqnarray}
where we introduce another numerical constant $c_2$ of order unity to make the evaluated relation equality.
Since we know $\omega=1$ for the Markovian limit $\alpha=1$, one of the numerical constants can be eliminated through
\begin{eqnarray}
c_2 = \left( \frac{3}{2}\right)^{1/2} c_1^{-1/2}
\end{eqnarray}
This leads to Eq.~(11) in main text with one fitting parameter $c_1$, which should be determined through the comparison with numerical simulation data.
As discussed in the main text, we found $c_1 = 0.1$ describes the simulation results well. The resultant dependence of $\omega$ on $\alpha$ is shown in Fig.~3(b) in the main text. Apparently, the relation is close to $\omega = \alpha$, but the value of $\omega$ is slightly larger than $\alpha$ in a systematic way. We note that, while irrelevant to the long time asymptotic power-law behavior, the short time behavior is highly sensitive to this $\omega$ exponent. For example, we show in Fig.~S1 the short time part of the FPT distribution $F(\tau)$ for the case of $\alpha=0.4$ and $0.5$, where our formula for $\omega (\alpha)$, but not $\omega = \alpha$, provides satisfactory fittings.

\vspace{0.2 cm}
\begin{figure}[ht]
	\centering
	\includegraphics[width=0.7\textwidth]{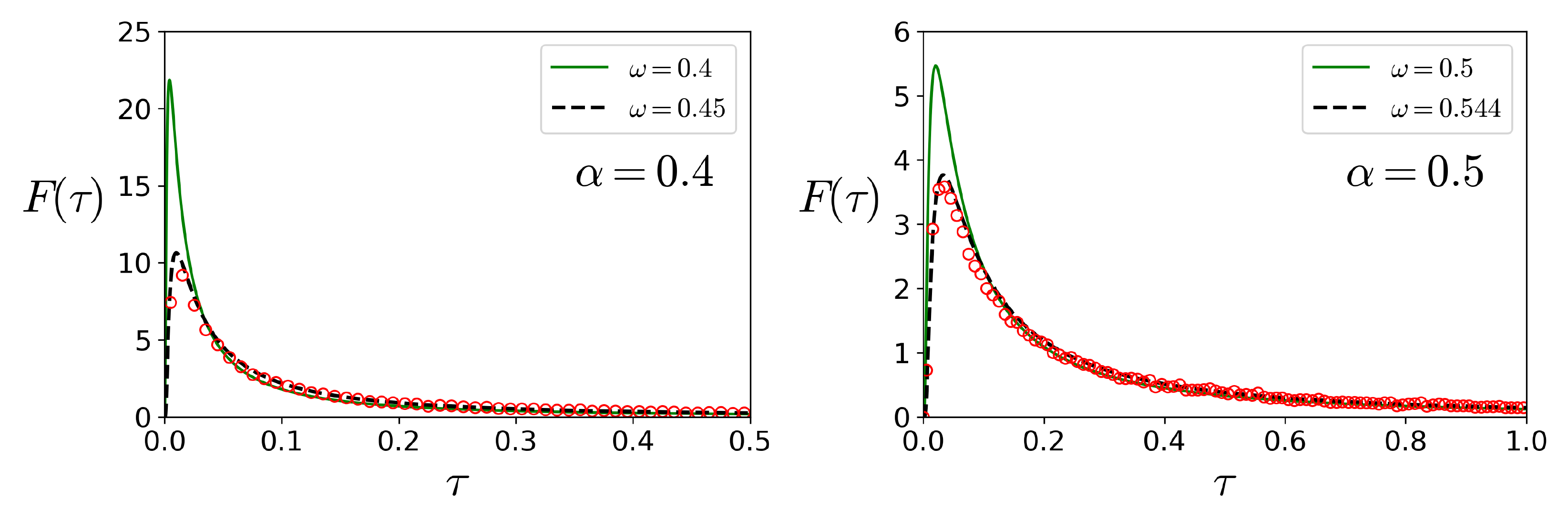}
	\caption{ {\bf Short time part of FPT distribution of non-Markovian walkers.} Plot of $F(\tau)$ for the case (a) $\alpha=0.4$ and (b) $\alpha=0.5$. The best fit values are $\omega =0.45$ for $\alpha=0.4$ and $\omega =  0.544$ for $\alpha=0.5$ , which are included in the plot of Fig.~3(b) in the main text.}
	\label{figS1}
			\vspace{0.4 cm}
\end{figure}

\section{Failure of the method of image}
The effect of the persistent memory in fBM becomes stronger with the departure from the Markovian limit $\alpha=1$. This is seen, for instance, in the spatial profile of $Q(x,t;1)$ shown in Fig. 4 in the main text, where the degree of the asymmetry $Q(x,t;1) \neq Q(-x,t;1)$, a hallmark of the memory effect, clearly shows up in $\alpha=0.5$ case, but less evident in $\alpha=0.8$ case. In such a situation, one may expect that the method of image, a standard method used in the Markovian system, might provide an acceptable approximate solution. In Fig.~S2, we show the probability of the survival walkers $P_{+}(x,t;1)$ for $\alpha=0.8, 0.5$ cases, where the comparison is made for our solution and that constructed by the method of image. Clearly, the method of image fails to capture the profile even qualitatively. In contrast, our method is capable of a quantitative description.  

\vspace{0.2 cm}
\begin{figure}[ht]
	\centering
	\includegraphics[width=0.7\textwidth]{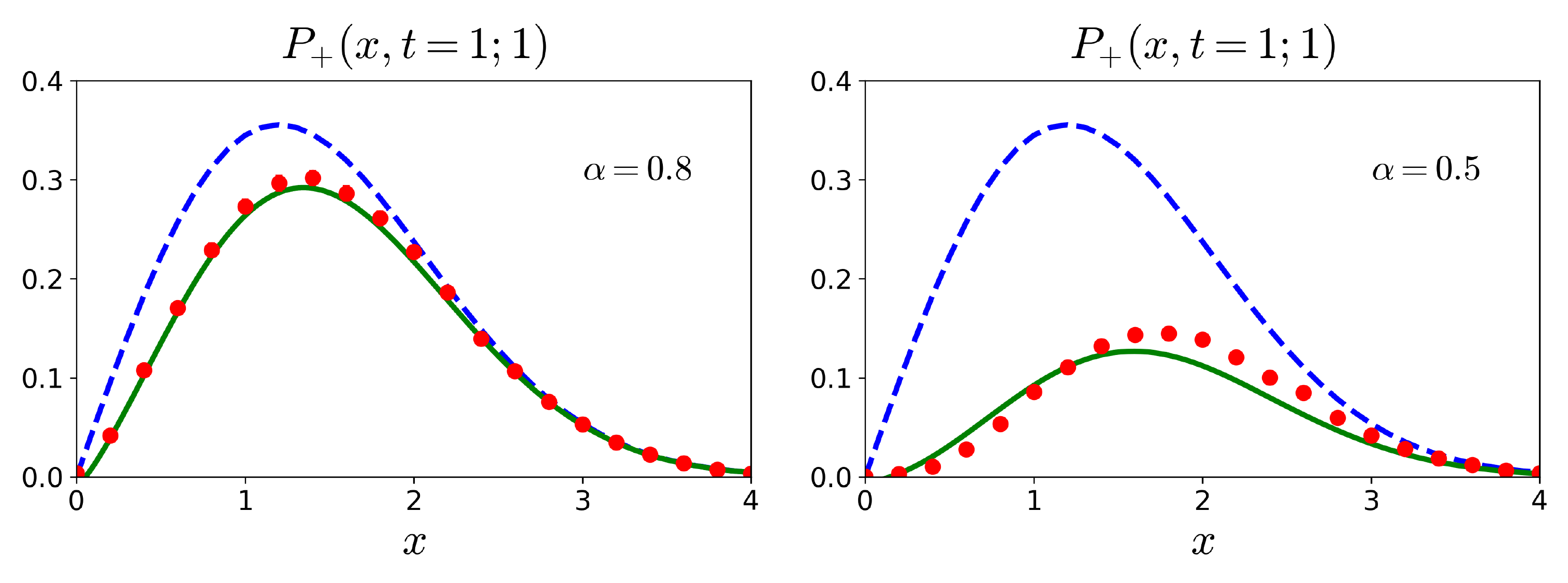}
	\caption{ {\bf Failure of the method of image.} Plot of $P_{+}(x,t;1)$ for (a) $\alpha=0.8$ and (b) $\alpha=0.5$. Solid curves are obtained from our theory, which quantitatively describe the numerical simulation result (symbols). In contrast, the method of image provide  qualitatively wrong profiles (dashed curves). }
	\label{figS2}
			\vspace{0.2 cm}
\end{figure}

\section{Numerical simulation}
To simulate fBM trajectories $\{x_0, x_1, \cdots, x_N \}$ of length $N$, we numerically integrated the discretized version of Eq.~(1) in main text with $f=0$.
The Gaussian variables $\eta_i$, called fractional Gaussian noise, have temporal correlation, whose long time part is characterized by the power-law memory as described after Eq.~(1) in main text.
To generate the fractional Gaussian noise, we employed the Davies and Harte algorithm~\cite{Davies_Harte_1987}, and generated $m$ samples of length $N$ for each $\alpha$. 
From these simulations, we calculated the standard deviation of the walker's displacement $\Delta x_N \equiv \sqrt{\langle (x_N - x_0)^2\rangle}$ after $N$ steps. To analyze the FPT statistics, we placed the hypothetical absorbing wall at $x = x_0 - {\tilde c} \, \Delta x_N$ such that the initial separation from the walker to the boundary is ${\tilde c} \, \Delta x$. We then reanalyzed each of $m$ trajectories to find its first arrival at the wall, and constructed the FPT distribution and the walkers' distribution after the FPT. We adopted $N=10^5$, $m=10^5$ and ${\tilde c}=1$ except for the FPT distribution data for long time regime (Fig. 2 (a) inset), where we adopted $N=10^6$ and $m=10^4$ and ${\tilde c}=0.5$.

\bibliography{stochastic}

